 \documentclass[twocolumn,epjc3]{svjour3}

\journalname{Eur. Phys. J. A}
 
\usepackage{epsfig}
\usepackage{color}
\usepackage{float}
\usepackage{amssymb}
\usepackage{amsmath, bbm}
\usepackage[figuresright]{rotating}

\usepackage[numbers,sort&compress]{natbib}
\usepackage{hyperref}

\newlength{\feynwidth} 
\usepackage{amssymb}
\usepackage{amsmath, bbm}
\usepackage[figuresright]{rotating}
\usepackage{capt-of}
\usepackage{multirow}

\def\nl{\nonumber\\}

\newcommand{\La}{{\Lambda}}
\newcommand{\Si}{{\Sigma}}

\newcommand{\be}{\begin{eqnarray}}
\newcommand{\ee}{\end{eqnarray}}

\newcommand{\eeb}{{e^+e^-}}
\newcommand{\LLb}{{\Lambda\bar\Lambda}}
\newcommand{\SSb}{{\Sigma\bar\Sigma}}
\newcommand{\NNb}{{N \bar N}}
\newcommand{\ppbar}{{p \bar p}}
\begin{document}

\title{{\boldmath$\Lambda\bar\Lambda$} final-state interaction in the reactions 
{\boldmath$e^+e^- \to \phi\La \bar \La$} and {\boldmath$e^+e^- \to \eta\La \bar \La$}}

\author{Johann Haidenbauer \thanksref{addr1,e2}
\and Ulf-G. Mei{\ss}ner \thanksref{addr2,addr1,e3}
}
\thankstext{e2}{e-mail: j.haidenbauer@fz-juelich.de}
\thankstext{e3}{e-mail: meissner@hiskp.uni-bonn.de}

\institute{Institute for Advanced Simulation (IAS-4), Institut f\"ur Kernphysik (IKP-3) and
  J\"ulich Center for Hadron Physics, Forschungszentrum J\"ulich, D-52428 J\"ulich, Germany
  \label{addr1}
           \and
           Helmholtz-Institut f\"ur Strahlen- und Kernphysik 
           and Bethe Center for Theoretical Physics,
           Universit\"at Bonn, D-53115 Bonn, Germany \label{addr2}
}

\date{\today}

\maketitle

\abstract{
Near-threshold
$\Lambda\bar\Lambda$ mass spectra for the reactions $e^+e^- \to \eta\Lambda\bar\Lambda$ 
and $e^+e^- \to \phi\Lambda\bar\Lambda$ are investigated with an emphasis on the role played 
by the interaction in the $\Lambda\bar\Lambda$ system. A variety of $\Lambda\bar\Lambda$ 
potential models is employed that have been established in the analysis of data on  
$p\bar p\to \Lambda\bar\Lambda$ in the past.
It is shown that the near-threshold enhancement observed for the two $e^+e^-$ reactions
can be reproduced by considering the $\Lambda\bar\Lambda$ final-state interaction
in the partial waves suggested by the helicity-angle analysis of the experiments. 
For $e^+e^- \to \eta\Lambda\bar\Lambda$ the same $\Lambda\bar\Lambda$ $S$-wave 
interaction as in $e^+e^- \to \Lambda\bar\Lambda$ is relevant and with it a 
consistent description of the pertinent measurements can be achieved. 
It is pointed out that a nonzero threshold cross section as observed for 
the latter reaction is not supported by the new $\eta\Lambda\bar\Lambda$ data.

}

\section{Introduction}

Experimental information on the properties of baryon-antibaryon ($B\bar B$) 
interactions is rather scarce, with the exception of the $\NNb$ system, of course, 
where standard scattering experiments are possible \cite{PS185}. For others, 
like those involving strange baryons and/or antibaryons, constraints on the 
forces can be only inferred from studies of reactions where the $B\bar B$ 
state is produced and interacts in the final state. 
Here, the by far best investigated system is $\LLb$. In particular, the hyperon
production process 
$\ppbar\to\LLb$ has been extensively measured in the PS185 experiment at the LEAR 
facility at CERN and data are available for total and differential cross sections, but also 
for spin-dependent observables \cite{PS185,PS1851,PS1852,PS1853,Barnes:2000,PS185:2006}, owing 
to the fact that one can exploit the self-analyzing weak $\Lambda$ decay. These data
span the energy range from the reaction threshold up to  $\sqrt{s}\approx 2.4$~GeV.

In the last two decades several other reactions producing a $\LLb$ pair have been
studied experimentally. Among them are the decays 
$B \to K\LLb$ \cite{Chang:2009}, $B \to D\LLb$ \cite{Lees:2014}, 
$J/\psi, \, \psi(3868) \to \eta\LLb, \pi^0\LLb$ \cite{BESIII:2013Psi,BESIII:2022Psi}, 
and $\psi(3868) \to \omega\LLb$ \cite{BESIII:2022o}. In addition, data 
for $\eeb \to \LLb$ \cite{Bisello:1990,Aubert:2007,Dobbs:2014,BESIII:2018,Ablikim:2019,BESIII:2021ee}, 
and for $\eeb \to \phi\LLb$ \cite{BESIII:2021}
and $\eeb \to \eta\LLb$ \cite{BESIII:2022} have been presented. 
Furthermore, there have been measurements of $\LLb$ correlation 
functions in heavy-ion collisions and in high energy $pp$ collisions by the ALICE 
Collaboration \cite{ALICE:2020,ALICE:2022}.
Finally, there are preliminary results from GlueX for $\gamma p \to p\LLb$ 
\cite{Li:2020,Pauli:2022}. See also the recent reviews \cite{Zhou:2022,Schonning:2023}. 

Among those reactions, the one that received the by far strongest attention by theorists 
is  $\eeb \to \LLb$, since it allows one to determine the electromagnetic 
form factors of the $\Lambda$ in the time-like region  
\cite{Baldini:2009,Dalkarov:2010,Haidenbauer:2016,Yang:2018,Cao:2018,Yang:2019,Xiao:2019,Ramalho:2020,Haidenbauer:2021,Li:2022,Lin:2022,Bystritskiy:2022,Tomasi-Gustafsson:2022}. 
One-photon exchange can be expected to dominate, so that the reaction mechanism is 
practically known, and the partial waves of the final state are restricted to either $^3S_1$ 
or $^3D_1$. 
Certainly the most conspicious aspect was the observation of a large non-zero cross section 
barely $1$~MeV away from the $\LLb$ threshold in the BES\-III experiment~\cite{BESIII:2018}. 
While the near-threshold energy dependence of the reaction cross section reported by 
the BaBar collaboration \cite{Aubert:2007} could be well described by assuming a 
standard final-state interaction (FSI) between the $\LLb$ pair \cite{Haidenbauer:2016},
the explanation of the very large ``at threshold'' BESIII value
required the inclusion of a so far unobserved narrow resonance 
\cite{Cao:2018,Li:2022}.

In the present work we re-investigate the FSI effects caused by the 
$\LLb$ interaction. This is done in view of the new precise measurements of the
near-threshold $\LLb$ invariant-mass spectrum by BESIII in the reactions 
$\eeb \to \phi\LLb$ \cite{BESIII:2021} and $\eeb \to \eta\LLb$ 
\cite{BESIII:2022}. The former process allows to examine the $\LLb$
interaction in other partial waves than in $\eeb\to\LLb$, see the
selection rules summarized in Table~\ref{tab:JPC}, while the latter
involves the same final state ($^3S_1$) so that one can explore whether 
the $\LLb$ FSI which describes the energy dependence of the $\eeb\to\LLb$ 
cross section can also explain the one in $\eeb \to \eta\LLb$.
In particular, it will be interesting to see whether the narrow structure 
required to describe the BESIII measurement of $\eeb\to\LLb$ \cite{Cao:2018,Li:2022} 
is also needed in and/or supported by the $\eeb \to \eta\LLb$ data.
  
As in our study of the reaction $\eeb\to\LLb$ \cite{Haidenbauer:2016}
we employ phenomenological $\LLb$ models that have been developed by the J\"ulich 
group for the analysis of $\ppbar \to \LLb$ data in the past
\cite{Haidenbauer:1992,Haidenbauer:1992A,Haidenbauer:1992B,Haidenbauer:1993}.
Indeed, in those studies several variants have been established which all describe the
PS185 data quite well. These variants have been already used by us for calculating the
$\eeb \to \LLb$ reaction and yielded cross sections well in line with the BaBar data. 
Moreover, it turned out that some of the interactions also
reproduce roughly the BESIII result \cite{Ablikim:2019} on the ratio and phase of the 
electromagnetic form factors $G_E$ and $G_M$ at $2.396$~GeV \cite{Haidenbauer:2021}. 

\begin{table}[t]
\renewcommand{\arraystretch}{1.5}
\begin{center}
\caption{Allowed $\LLb$ partial waves and $J^{PC}$ assignments 
(up to $P$-waves) for various channels in the reaction $\eeb\to\gamma\to X$.}
\begin{tabular}{l|c}
\hline
final state & partial waves \\
\hline 
$\phi\LLb$ & ${}^1S_0 \, [0^{-+}], {}^3P_0 \, [0^{++}], {}^3P_1 \, [1^{++}], {}^3P_2 \, [2^{++}]$  \\
$\eta\LLb$ & ${}^3S_1 \, [1^{--}], {}^1P_1 \, [1^{+-}]$  \\
$\LLb$ & ${}^3S_1 \, [1^{--}]$  \\
\hline
\end{tabular}
\label{tab:JPC}
\end{center}
\renewcommand{\arraystretch}{1.0}
\end{table}

The paper is structured in the following way. In the subsequent section we
provide a brief overview of the employed formalism for treating the FSI effects. 
In Sect.~\ref{sec:Results} we present our results. Specifically, we review
the situation for $\eeb\to\LLb$ and then show our predictions for the 
$\LLb$ invariant mass spectra measured in the reactions 
$\eeb \to \eta\LLb$ and $\eeb \to \phi\LLb$. We also discuss the
situation for $\psi (3868) \to \eta\LLb$ and $\gamma p \to p\LLb$.  
The paper closes with a short summary.  The appendix contains a brief discussion
of the $\LLb$ correlations measured by ALICE.


\section{Treatment of the $\LLb$ final-state interaction}
\label{Sec:Form}

Our calculation of the $\LLb$ invariant-mass spectrum is based on the distorted wave Born 
approximation (DWBA), where the reaction amplitude $A$ is given schematically by 
\cite{SibirtsevPRD,Kang:2015}
\begin{equation}
A=A^0 + A^0 G^{\LLb} T^{\LLb} \ . 
\label{eq:DWBA}
\end{equation}
Here, $A^0$ is the elementary (or primary) production amplitude, $G^{\LLb}$ the free $\LLb$ Green's
function, and $T^{\LLb}$ the $\LLb$ reaction amplitude.
For a particular (uncoupled) $\LLb$ partial wave with orbital angular momentum $L$ Eq.~(\ref{eq:DWBA}) 
reads
\begin{equation}
A_L=A_L^{0}+\int_0^\infty \frac{dp p^2}{(2\pi)^3} A_L^{0} 
\frac{1}{2E_k-2E_p+i0^+}T_{L}(p,k;E_k) ,
\label{eq:dwba1}
\end{equation}
where $T_{L}$ denotes the partial-wave projected $T$-matrix, and $k$ and $E_k$ are
the momentum and energy of the $\Lambda$ (or $\bar\Lambda$) in the center-of-mass system of the 
$\LLb$ pair.
The quantity $T_{L}(p,k;E_k)$ is obtained by solving the Lippmann-Schwinger (LS) equation,

\begin{eqnarray}
&&T_{L}(p',k;E_k)=V_{L}(p',k)+ \nl
&& \int_0^\infty \frac{dpp^2}{(2\pi)^3} \, V_{L}(p',p)
\frac{1}{2E_k-2E_p+i0^+}T_{L}(p,k;E_k)~,\nl 
\label{LS}
\end{eqnarray}
for a specific $\LLb$ potential $V_L$. In the case of coupled partial waves like the $^3S_1$--$^3D_1$
the corresponding coupled LS equation is solved \cite{Haidenbauer:2014}, and then 
$T_{LL}$ is used in Eq.~(\ref{eq:dwba1}).

In principle, the elementary production amplitude $A^0_L$ in Eq.~\eqref{eq:dwba1} 
depends on the total energy and also on the $\LLb$ momentum and the momentum of the
additional particle relative to the $\LLb$ system \cite{Gasparyan}.  
In the near-threshold region $A^0_L = \bar A^0_L k^L$. 
The variation of $\bar A^0_L$ with regard to the other variables should be rather 
small as compared to the strong momentum dependence induced by the $\LLb$ FSI. 
Therefore, we neglect it in the following so that Eq.~\eqref{eq:dwba1} reduces to

\begin{eqnarray}
&&A_L=\bar A^0_L k^L \times \nonumber \\ 
&&\left[1+ \int_0^\infty \frac{dp p^2}{(2\pi)^3} 
\frac{p^L}{k^L}\frac{1}{2E_k-2E_p+i0^+}T_{L}(p,k;E_k)\right] .\nl 
\label{eq:dwba2}
\end{eqnarray}
%
with $\bar A^0_L$  a constant. We found, however, that a straight and simplistic 
off-shell extension $\bar A^0_L k^L \to \bar A^0_L p^L$ in the integral in 
Eq.~(\ref{eq:dwba2}) leads to a strong and seemingly artificial enhancement of the 
principal-value part for $P$-waves, see also the discussion in Ref.~\cite{Entem:2007}.  
Therefore, in order to avoid this artifact we attenuate the $P$-wave momentum factor 
$p$ ($k$) by replacing it with $p \exp{(-p^2/\Lambda^2)}$ (and likewise for $k$) in 
the actual calculation. 
The considered value for $\Lambda$ ($500 \sim 600$~MeV/c) are chosen in line with
our experience in studying FSI effects with chiral $N\bar N$ potentials \cite{Kang:2015} 
where the cutoff of the intrinsic regulator is of that order. Note that
$k\simeq \Lambda \simeq 500$~MeV/c corresponds to an invariant mass of roughly $2.45$~GeV. 
Thus, with such a value the on-shell properties of the $\LLb$ amplitude in 
the region near the $\LLb$ threshold we are interested in remain practically unmodified. 
For a thorough discussion of various aspects of the treatment of FSI effects, 
see Refs.~\cite{Gasparyan,Hanhart,Baru} and also \cite{Watsonbook}.

The $\LLb$ invariant-mass spectrum is calculated via 
\begin{equation}
\frac{d\sigma}{dM (\LLb )} \propto k\,|A_L(k)|^2 \ ,
\label{eq:INV}
\end{equation}
which is valid when the (total) energy is significantly larger than the $\phi\LLb$ or 
$\eta\LLb$ threshold energies. Then for low $\LLb$ invariant masses the relative 
momentum of the third particle is large and it does not distort the signal of interest. 
This condition is well fulfilled by the BESIII measurements.   

In the present calculations we employ $\LLb$ potentials which were established within the 
J\"ulich meson-baryon model in studies of the reaction $\ppbar \rightarrow \LLb$ 
\cite{Haidenbauer:1992,Haidenbauer:1992A,Haidenbauer:1992B,Haidenbauer:1993}. In those 
works the hyperon-production reaction is considered within a coupled-channel approach, 
which allowed to take into account rigorously the effects of the initial ($\ppbar$) and 
final ($\LLb$) state interactions. Both play a role for describing the 
data for energies near the production threshold. For details of the potentials
we refer the reader to the cited works. Here we just want to mention that 
the elastic parts of the interactions in the $\ppbar$ and $\LLb$ channels 
are described by meson exchanges, whereas annihilation processes are accounted 
for by phenomenological optical potentials. 
 
We consider a variety of potentials in order to assess a possible (and unavoidable) model 
dependence of the results. Specifically, we use the $\LLb$ potentials I, II, and III of
Ref.~\cite{Haidenbauer:1992} (cf. Table~III) and ``K'' from
Ref.~\cite{Haidenbauer:1992B} (Table~II), denoted by IV below. 
The models differ by variations in the employed parameterization of the $N\bar N$ 
annihilation potential and by differences in the $\ppbar \rightarrow \LLb$ 
transition mechanism.
All of them provide a rather good overall description of the wealth of
$\ppbar \to \LLb$ data collected by the P185 Collaboration~\cite{PS185}.
In particular, the total reaction cross sections produced by those 
potentials agree with each other and with the experiment up to
$p_{\rm lab} \approx 1700$~MeV/c (corresponding to $\sqrt{s}\approx 2.32$~GeV
or an excess energy $Q=\sqrt{s}-2m_\Lambda$ of about $90$~MeV). 
Even spin-dependent observables (analyzing powers, spin-correlation parameters) 
are, in general, described fairly well. 
For reference, in the appendix the scattering lengths of those potentials are 
summarized and a brief discussion of results for the $\LLb$ correlations measured 
by the ALICE Collaboration \cite{ALICE:2022} is provided. 
Finally, we emphasize again that the validity of treating FSI effects via 
Eqs.~(\ref{eq:dwba2}) and (\ref{eq:INV}) is clearly limited, say to excess energies 
of $50$ to $100$~MeV. 
With increasing invariant mass the momentum dependence of the reaction/production 
mechanism should become more and more relevant and will likewise influence the 
invariant-mass spectrum. Last but not least when approaching the $\SSb$ threshold 
the overall dynamics could change significantly \cite{Haidenbauer:2016}.  


\section{Results}
\label{sec:Results} 

In the following we present predictions for the $\LLb$ inva\-riant-mass spectrum
for the reactions $\eeb\to\eta\LLb$ and $\eeb\to\phi\LLb$. The former is 
interesting because it involves the same $\LLb$ state (FSI) as 
$\eeb\to\LLb$. In the latter case FSI effects in $\LLb$ $P$-waves are
expected to play a role, considering the helicity-angle analysis of the 
BESIII data \cite{BESIII:2021}. 
As mentioned above, the calculations utilize $\LLb$ potentials 
from \cite{Haidenbauer:1992,Haidenbauer:1992B} which have been already
explored in our study of the electromagnetic form factors of the $\Lambda$ 
in the time-like region \cite{Haidenbauer:2016}. 

\begin{figure}
\begin{center}
\includegraphics[height=88mm,angle=-90]{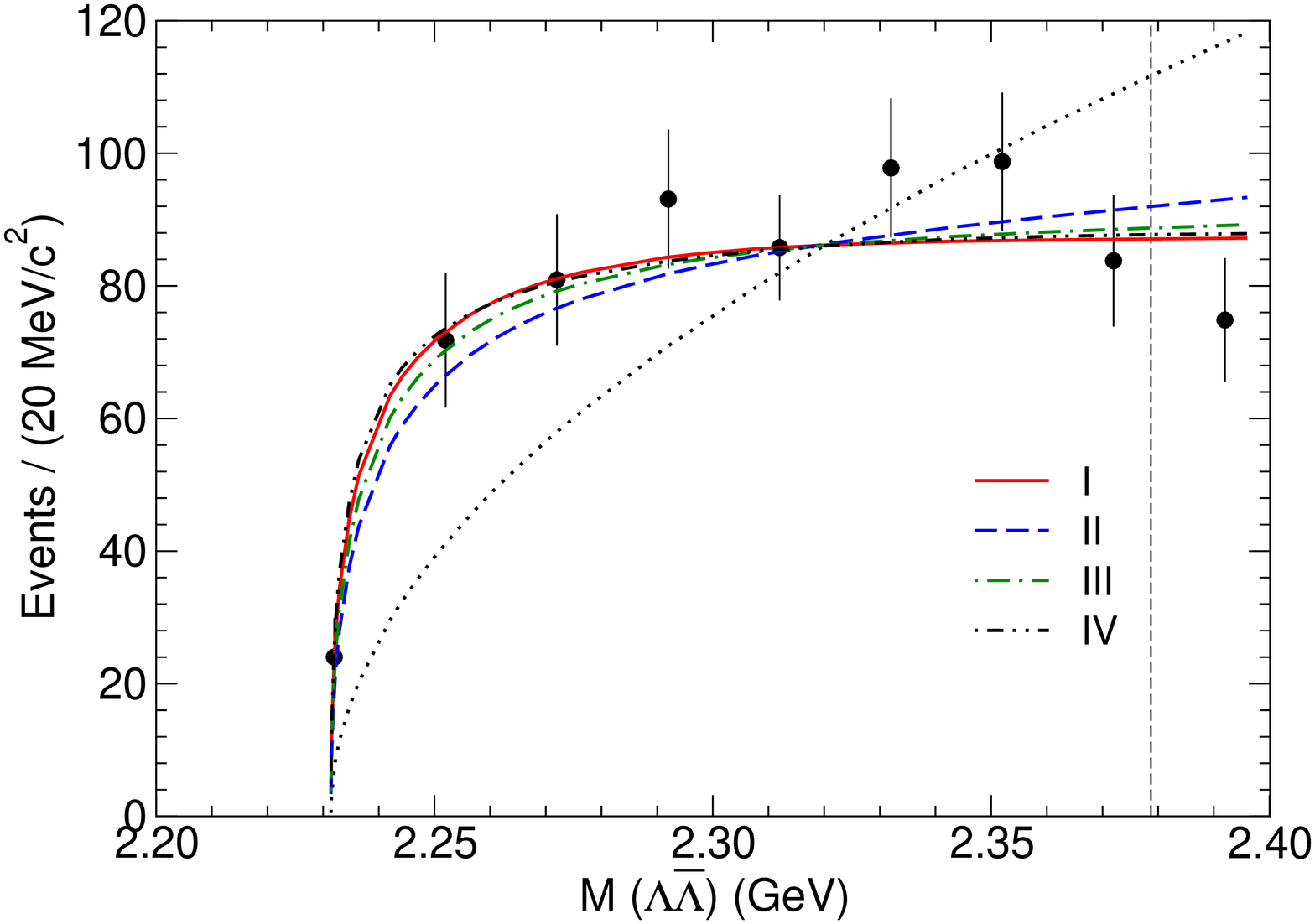}

\includegraphics[height=88mm,angle=-90]{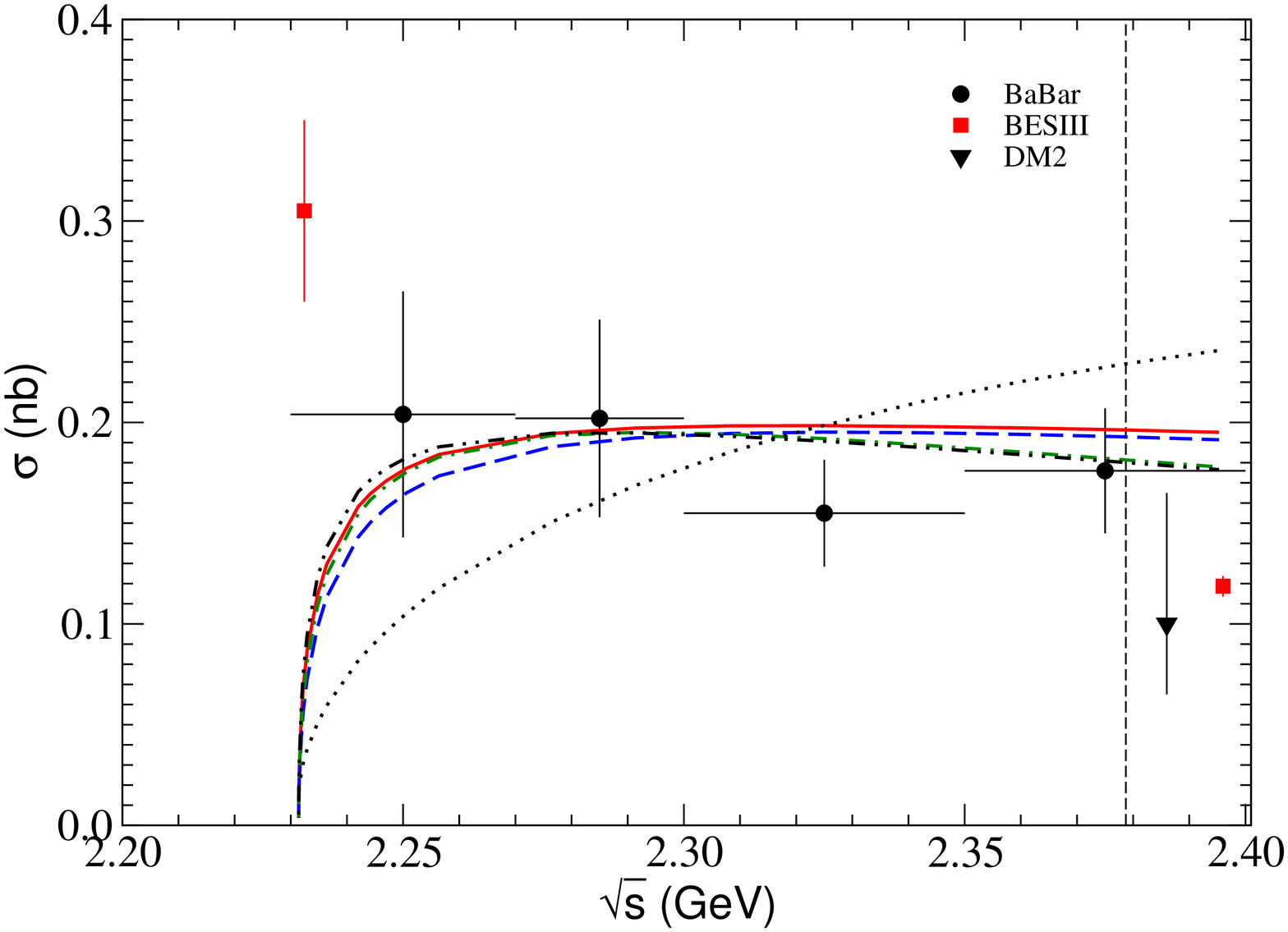}
\caption{Top: Results for $\eeb\to \eta \LLb$ based on the $^3S_1$ partial wave 
of the $\LLb$ models I-IV \cite{Haidenbauer:1992,Haidenbauer:1992B}. 
Data are from Ref.~\cite{BESIII:2022}. 
The phase space behavior is indicated by the dotted line. The vertical
thin dash-dotted line marks the $\Si^+\overline{\Si^+}$ threshold which
is around $2.379$~GeV. All curves are arbitrarily normalized so that they 
coincide at $\sqrt{s}\approx 2.32$~GeV. 
Bottom: Corresponding results for $\eeb \to \LLb$, see Refs.~\cite{Haidenbauer:2016,Haidenbauer:2021}.
Data are from Refs.~\cite{Bisello:1990} (DM2), \cite{Aubert:2007} (BaBar), and 
\cite{BESIII:2018} (BESIII). 
}
\label{fig:lle}
\end{center}
\end{figure}


\subsection{The reaction $\eeb\to\eta\LLb$} 

We start with reminding the reader on the situation for 
$\eeb\to\LLb$. Experimental results in the near-threshold 
region are presented in Fig.~\ref{fig:lle} (bottom), together
with our predictions based on various $\LLb$ potentials
\cite{Haidenbauer:2016,Haidenbauer:2021}. One can see on the one hand that all 
interactions yield a more or less flat behavior of the cross section, 
in line with the BaBar data \cite{Aubert:2007}. 
On the other hand, the data point from BESIII \cite{BESIII:2018}
at $2232.4$~MeV, i.e. just about $1$~MeV above the $\LLb$ threshold at
$2m_\Lambda = 2231.4$~MeV, deviates from the overall trend and is not reproduced
by our calculation. As demonstrated in works 
by others, that data point can be only described by introducing a narrow 
and so far unknown resonance located very near the threshold
($M_x = 2230.9$~MeV, $\Gamma_x = 4.7$~MeV) \cite{Li:2022} or by a 
suitably adjusted contribution of the sub-threshold resonance $\phi(2170)$  
in combination with an additional resonance at $2340$~MeV \cite{Cao:2018}.

\begin{figure}
\begin{center}
\includegraphics[height=88mm,angle=-90]{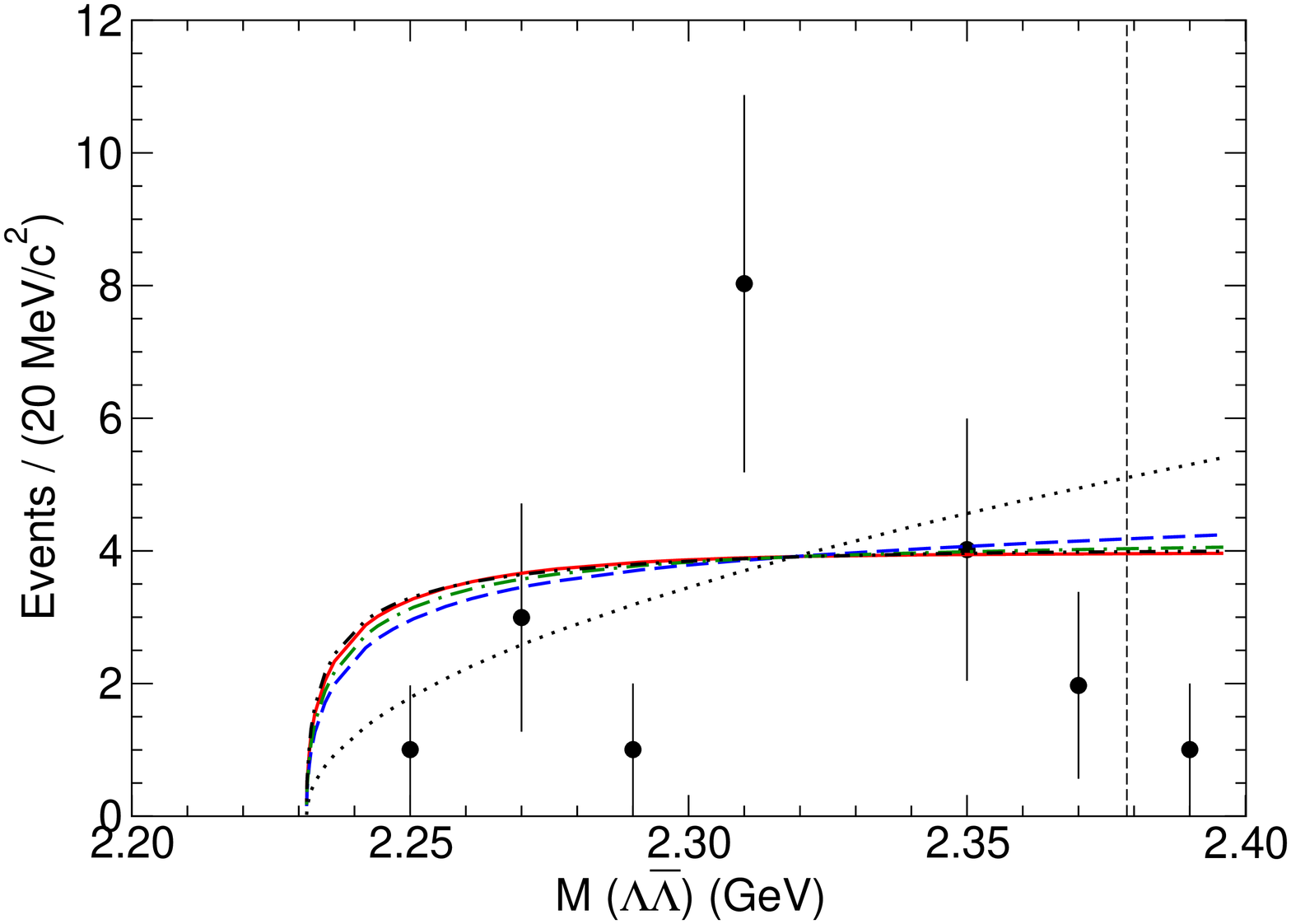}

\includegraphics[height=88mm,angle=-90]{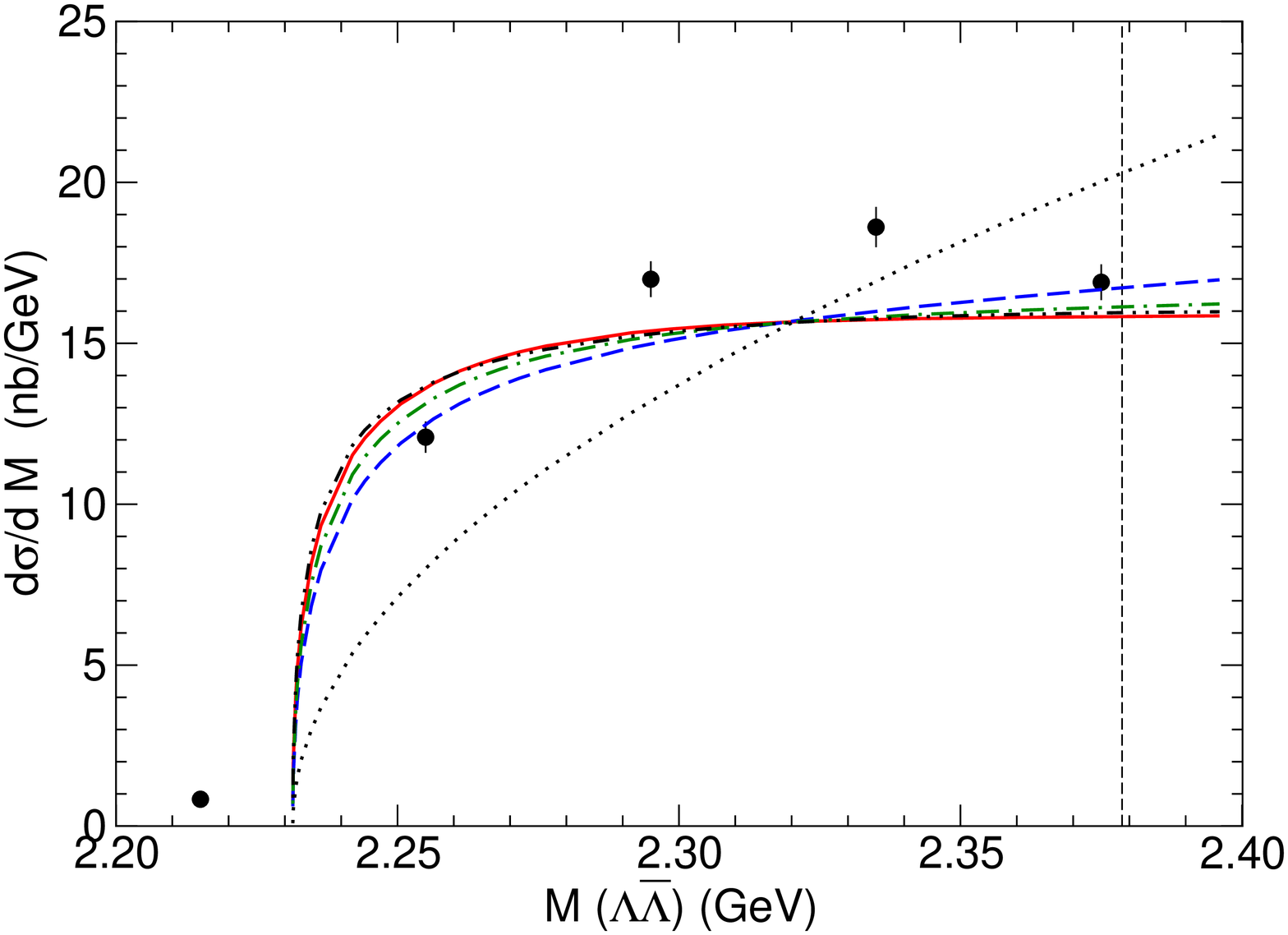}
\caption{Invariant-mass spectra for $\psi(3686) \to \eta\LLb$ 
from the BESIII Collaboration \cite{BESIII:2022Psi} (top) and preliminary 
results for $\gamma p \to p\LLb$ from GlueX \cite{Pauli:2022} (bottom). 
Same description of curves as in Fig~\ref{fig:lle}. 
}
\label{fig:gluex}
\end{center}
\end{figure}

Interestingly, initial data for $\ppbar \to \LLb$ provided some support for  
a near-threshold resonance in the $^3S_1$ $\LLb$ state \cite{Carbonell:1993}.
However, a later high-statistics measurement by Barnes et al.~\cite{Barnes:2000}
ruled that out. In fact, including a narrow near-threshold resonance in our $\LLb$ 
potentials would completely spoil the agreement with the $\ppbar \to \LLb$ data.
It should be mentioned that a partial-wave analysis of the reaction $\ppbar \to \LLb$, 
performed by Bugg \cite{Bugg:2004} indicated again the presence of a resonance. 
However, its width is very large ($\Gamma \approx 275$~MeV). In our opinion that
resonance has to be considered simply as an effective parameterization of the 
FSI effects in the $^3S_1$-$^3D_1$ partial wave rather than as indication for 
a physical state. 

The new measurement of the $\LLb$ mass spectrum in the 
reaction $\eeb\to\eta\LLb$ is shown in Fig.~\ref{fig:lle} (top).
The possible $\LLb$ states near threshold are 
$^3S_1$ ($1^{--}$) and $^1P_1$ ($1^{+-}$), where the former 
is the partial wave which also causes the enhancement in the 
$\eeb\to\LLb$ cross section. Indeed, the invariant-mass spectrum 
calculated via Eq.~(\ref{eq:INV}), including FSI effects from our 
$\LLb$ potentials in the $^3S_1$ partial wave, describes the BESIII 
data strikingly well over the whole considered invariant-mass region. 
Note specifically the one data point very close to threshold 
which is exactly in line with the energy dependence generated 
by the FSI. In view of that, it is hard to believe that something 
unusual happens only in $\eeb\to\LLb$ at the threshold whereas for 
other reactions a consistent and convincing description of the experiments 
is achieved by the FSI in the same $\LLb$ partial wave. 

In Fig.~\ref{fig:gluex} we show result for two other reactions where 
the $\LLb$ invariant-mass spectrum has been measured. Also in those 
cases we expect that the $^3S_1$ partial wave provides the dominant 
FSI effect. 
Unfortunately, the statistics of the BESIII data on $\psi (3686)\to\eta\LLb$
(top) is too low for allowing reliable conclusions.
The invariant-mass spectrum from a measurement of the reaction
$\gamma p\to p\LLb$ by GlueX is roughly in line with our prediction
based on the FSI by the $^3S_1$ partial wave, but we want to emphasize 
that those data are still preliminary. Moreover, for that reaction
there are no strict selection rules so that the $^1S_0$
partial wave can likewise contribute and FSI effects there could have 
an impact on the result, too. 

\begin{figure}
\begin{center}
\includegraphics[height=88mm,angle=-90]{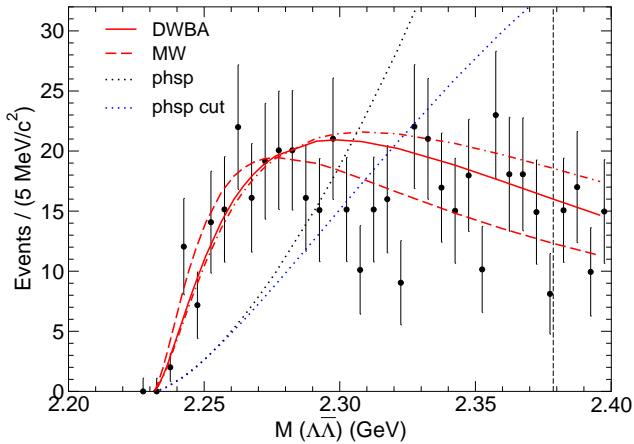}
\caption{Invariant-mass spectrum for $\eeb\to \phi \LLb$. 
The results are based on the $^3P_2$ partial wave of the $\LLb$
model I \cite{Haidenbauer:1992}.  
The curves correspond to the DWBA calculation (with cutoff $500$ (solid) 
and $600$~MeV (dash-dotted) and the Migdal-Watson approximation (dashed), 
see text. The $P$-wave phase space ($k^3$), without (phsp) and with 
cutoff (phsp cut), is indicated by the dotted lines.  
Data are from Ref.~\cite{BESIII:2021}.
}
\label{fig:llpT}
\end{center}
\end{figure}


\subsection{The reaction $\eeb\to\phi\LLb$} 

Recently, the BESIII Collaboration \cite{BESIII:2021} has also 
published data for the reaction $\eeb\to\phi\LLb$ \cite{BESIII:2021}.
The measurement is characterized by an excellent momentum resolution
and by the fact that the angular distributions have been measured
and analysed. According to that analysis the $\LLb$ pair is 
produced in the $1^{++}$, $2^{++}$, or $2^{-+}$ states, i.e. in the
$\LLb$ partial waves $^3P_1$, $^3P_2$, or $^1D_2$. For illustration, 
in the following we will show results for all triplet $P$-waves
($^3P_0$, $^3P_1$, $^3P_2$) of the $\LLb$ potentials. 

\begin{figure}
\begin{center}
\includegraphics[height=88mm,angle=-90]{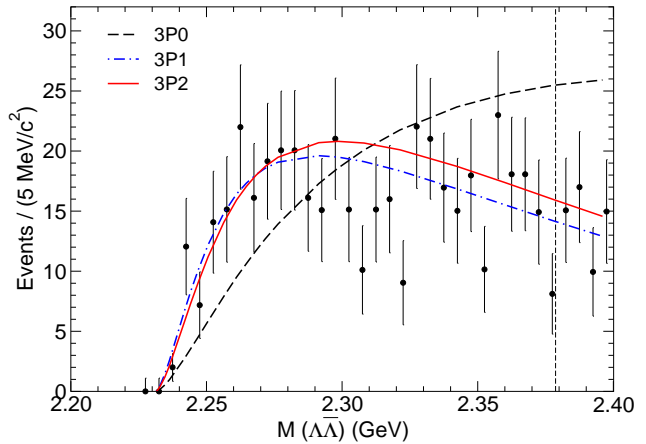}

\includegraphics[height=88mm,angle=-90]{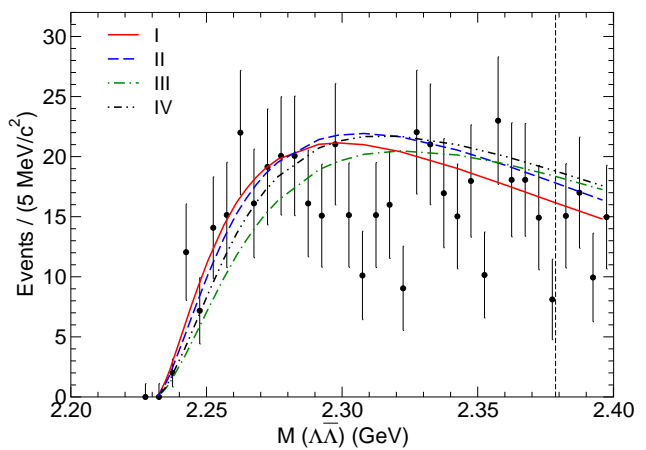}
\caption{Invariant-mass spectrum for $\eeb\to \phi \LLb$.
Top: Results for different $\LLb$ partial waves ($^3P_0$, $^3P_1$, $^3P_2$) 
based on $\LLb$ model I \cite{Haidenbauer:1992}. 
Bottom: Results for different $\LLb$ models I, II, III \cite{Haidenbauer:1992},
and IV \cite{Haidenbauer:1992B}, based on the $^3P_2$ partial wave. 
Data are from Ref.~\cite{BESIII:2021}.
}
\label{fig:llp}
\end{center}
\end{figure}

However, first we focus on aspects of the treatment of $P$-wave interactions in 
our FSI formalism.  As already mentioned in Sect.~\ref{Sec:Form}, 
in the case of $P$-wave interactions we include a cutoff in the evaluation
within the DWBA (\ref{eq:dwba2}). In Fig.~\ref{fig:llpT} we examine the 
effect of this treatment. The (upper) dotted line represents the phase-space
behavior of a $P$-wave, where the invariant-mass spectrum 
is then proportional to $k^3$. In case of the blue (lower) dotted line the cutoff
factor is multiplied. One can see that the change is only small
to moderate over the interesting region of say $M(\LLb) \lesssim 2.32$~GeV. 
Specifically, no additional and undesirable energy dependence is introduced.
The effect of the FSI, calculated via Eq.~(\ref{eq:dwba2}), leads
to a drastic modification of the spectrum, visualized here for the $^3P_2$ partial wave 
of model~I. For illustration we show predictions based on a cutoff of $500$~MeV (solid line) and
for $600$~MeV (dash-dotted line), and also results obtained within the so-called 
Migdal-Watson (MW) approximation (dashed line) where the invariant-mass spectrum 
is basically given by $T_L(k,k;E_k)/k$ \cite{Entem:2007}. 
One can see that all the results are qualitatively similar. 

Predictions for the invariant-mass spectrum based on the $^3P_0$, $^3P_1$, 
and $^3P_2$ partial waves of the $\LLb$ potential~I are shown in 
Fig.~\ref{fig:llp} (top). Here $500$~MeV is used for the cutoff. 
Obviously, the result for the $^3P_1$ as well as those for the $^3P_2$
are well in line with the experiment. Those two states were also 
favored by the helicity-angle analysis of BESIII \cite{BESIII:2021}. 
For the $^3P_0$ partial wave, which anyway is practically excluded by 
that analysis, the FSI effects would fall short to describe the data. 
In order to complete the picture, in Fig.~\ref{fig:llp} (bottom)
we show results based on all $\LLb$ potentials~I-IV, selectively for 
the $^3P_2$ partial wave. Obviously, the model-dependence of the 
predictions is quite small, something we already observed above
when considering FSI effects due to the $^3S_1$ state.

The invariant-mass spectrum for $\eeb\to\phi\LLb$ was also studied by
Milstein and Salnikov \cite{Milstein:2022} and they too achieved agreement 
with the data by including FSI effects. They also showed that with 
the $^1D_2$ partial wave (i.e. the $2^{-+}$ state) one cannot describe 
the near-threshold behavior and, therefore, we did not consider this state 
in our calculations. Anyway, we want to emphasize that in their investigation 
the potential was specifically constructed for and fitted to the BESIII data. 
As said above, the $\LLb$ potentials employed by us were established in a study 
of the $\ppbar \to \LLb$ reaction and fitted to 
the LEAR data. Thus our result for the $\LLb$ invariant-mass spectrum are in fact 
predictions. We note that our potentials are also more realistic because they 
include effects from $\LLb$ annihilation \cite{Haidenbauer:1992}.

\section{Summary}

In the present work we have investigated invariant-mass spectra for the reactions 
$\eeb  \to \eta\LLb$ and $\eeb \to \phi\LLb$ close to the $\LLb$ threshold.
Specific emphasis has been put on the role played by the interaction in the final
$\LLb$ system which is taken into account rigorously. For it  
a variety of $\LLb$ potential models have been employed that 
have been established in the analysis of data on the reaction 
$\ppbar\to \LLb$ from the LEAR facility at CERN. 
 
It turned out that the near-threshold invariant-mass dependence of the $\LLb$ 
spectra observed in those two reactions can be well reproduced by considering 
the $\LLb$ FSI in the partial waves suggested by the helicity-angle analysis of 
the experiment. In the case of $\eeb \to \eta\LLb$ the partial wave responsible
for the FSI ($^3S_1$) is identical to the one which dominates the $\eeb\to \LLb$ 
cross section near threshold. It is shown that the enhancement generated by the FSI 
in this state allows one to achieve a consistent description of both reations.
However, a nonzero ``threshold'' cross section as suggested for the latter 
reaction in the BESIII experiment \cite{BESIII:2018} is not observed in the 
new $\eta\LLb$ data.
In fact, none of the reactions with $\LLb$ in the final state,
measured in recent times, confirms or supports the existence of a
narrow near-threshold resonance that couples to the $\LLb$ system. 

\vspace{0.5cm}
\noindent
{\bf Acknowledgements:}
Work supported by the European Research Council (ERC) under the European
Uni\-on's Horizon 2020 research and innovation programme
(grant no.~101018170, EXOTIC), and by the DFG and the NSFC through
funds provided to the Sino-German CRC 110 ``Symmetries and
the Emergence of Structure in QCD'' (Project number 196253076 - TRR~110).
The work of UGM was also supported by the Chinese Academy of Sciences (CAS)
through a President's International Fellowship Initiative (PIFI)
(Grant No. 2018DM0034) and by the VolkswagenStiftung
(Grant No. 93562).

\appendix 
\section{Predictions for the $\LLb$ correlation function}

For completeness, in Table \ref{tab:ere} we summarize the scattering lengths
for the employed $\LLb$ potentials. 

\begin{table}[htbp]
\renewcommand{\arraystretch}{1.5}
\begin{center}
\caption{$\LLb$ scattering lengths (in fm) in the $^1S_0$ and $^3S_1$ partial waves
 of the employed $\LLb$ potentials \cite{Haidenbauer:1992,Haidenbauer:1992B}. 
The spin-averaged value by the ALICE Collaboration is from an 
analysis of the $\LLb$ correlation function measured in Pb-Pb collisions
\cite{ALICE:2020}.}
\begin{tabular}{l|c|c}
\hline
potential & $a (^1S_0)$ & $a (^3S_1)$ \\
\hline 
I   &   $0.32 -{\rm i}  0.52$  & $ 0.74 -{\rm i}  0.56$ \\
II  &   $0.67 -{\rm i}  1.14$  & $ 0.66 -{\rm i}  0.37$ \\
III &   $1.42 -{\rm i}  1.15$  & $ 1.00 -{\rm i}  0.44$ \\
IV  &   $1.56 -{\rm i}  1.40$  & $ 0.98 -{\rm i}  0.65$ \\
\hline
ALICE & \multicolumn{2}{c}{$(0.90\pm 0.16)-{\rm i}(0.40\pm 0.18)$} \\
\hline
\end{tabular}
\label{tab:ere}
\end{center}
\renewcommand{\arraystretch}{1.0}
\end{table}

\begin{figure}
\begin{center}
\includegraphics[height=88mm,angle=-90]{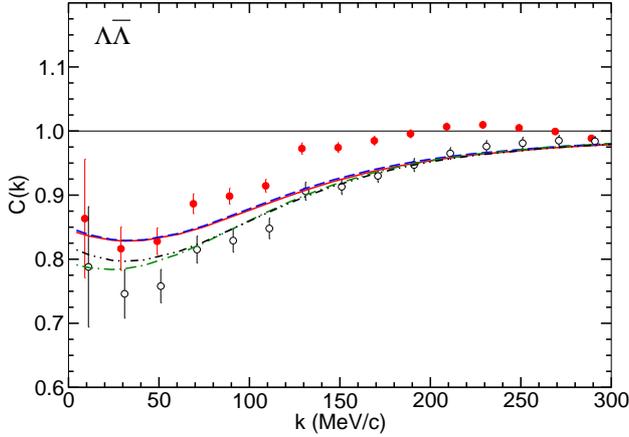}
\caption{$\LLb$ correlation function measured in $pp$ collisions at
$13$~TeV by the ALICE Collaboration \cite{ALICE:2022}. Filled symbols
are the original data while the opaque symbols include corrections 
for the background as estimated in that work.  
The calculation is based on the $\LLb$ wave functions in the
$^1S_0$ and $^3S_1$ partial waves, see text. 
Same description of curves as in Fig.~\ref{fig:lle}. 
}
\label{fig:femto}
\end{center}
\end{figure}

Furthermore, 
for illustration we provide predictions for the two-particle momentum 
correlation function in comparison to data of the ALICE Collaboration from
$pp$ collisions at $\sqrt{s}=13$~TeV \cite{ALICE:2022}, see Fig.~\ref{fig:femto}. 
We want to emphasize that these results are only meant for providing a qualitative 
impression. The calculations were performed with the wave functions in the $^1S_0$ 
and $^3S_1$ partial waves in the standard way, assuming a Gaussian function for the 
source \cite{Cho:2017,Haidenbauer:2018LL,Haidenbauer:2022LL}. However, possible contributions 
from higher partial waves and from the annihilation channels were omitted, see 
\cite{Haidenbauer:2018LL,ALICE:2022} for more details. In addition, no 
adjustment of femtoscopic parameters, like the source radius $R$ and the so-called
feed-down parameter $\lambda$ \cite{Cho:2017}, was done. Here we simply chose  
values ($R=1.1$~fm, $\lambda=0.35$) comparable to those suggested in \cite{ALICE:2022}.
Nonetheless, one can see that there is a good qualitative agreement with the
measurement for all the four potentials. Remarkably, the moderate rise of the
correlation function at very low momenta indicated by the data is reproduced 
by the calculations. In any case, there is no indication for a near-threshold 
resonance. 

\begin{thebibliography}{99}

\bibitem{PS185} 
E.~Klempt, F.~Bradamante, A.~Martin and J.~M.~Richard,
Phys. Rept. \textbf{368}, 119 (2002). 

\bibitem{PS1851} P.~D. Barnes {\it et al.},
  Phys. Lett. B {\bf 229}, 432 (1989).
\bibitem{PS1852} P.~D. Barnes {\it et al.},
  Nucl. Phys. A {\bf 526}, 575 (1991).
\bibitem{PS1853} P.~D. Barnes {\it et al.},
  Phys. Rev. C {\bf 54}, 1877 (1996).
\bibitem{Barnes:2000}
  P.~D.~Barnes {\it et al.},
  Phys.\ Rev.\ C {\bf 62}, 055203 (2000).
\bibitem{PS185:2006}
K.~D.~Paschke \textit{et al.},
Phys. Rev. C \textbf{74}, 015206 (2006).


\bibitem{Chang:2009}                   
Y.~W.~Chang \textit{et al.} [Belle],
Phys. Rev. D \textbf{79}, 052006 (2009).

\bibitem{Lees:2014}
J.~P.~Lees \textit{et al.} [BaBar],
Phys. Rev. D \textbf{89}, 112002 (2014).


\bibitem{BESIII:2013Psi}
M.~Ablikim \textit{et al.} [BESIII],
Phys. Rev. D \textbf{87}, 052007 (2013). 

\bibitem{BESIII:2022Psi}
M.~Ablikim \textit{et al.} [BESIII],
Phys. Rev. D \textbf{106}, 072006 (2022). 

\bibitem{BESIII:2022o}
M.~Ablikim \textit{et al.} [BESIII],
Phys. Rev. D \textbf{106}, 112011 (2022).

\bibitem{Bisello:1990}
  D.~Bisello {\it et al.} [DM2 Collaboration],
  Z.\ Phys.\ C {\bf 48}, 23 (1990).

\bibitem{Aubert:2007}
  B.~Aubert {\it et al.} [BaBar Collaboration],
  Phys.\ Rev.\ D {\bf 76}, 092006 (2007).

\bibitem{Dobbs:2014}
S.~Dobbs, A.~Tomaradze, T.~Xiao, K.~K.~Seth and G.~Bonvicini,
Phys. Lett. B \textbf{739}, 90 (2014).

\bibitem{BESIII:2018}
M.~Ablikim \textit{et al.} [BESIII],
Phys. Rev. D \textbf{97}, 032013 (2018).

\bibitem{Ablikim:2019}
M.~Ablikim \textit{et al.} [BESIII],
Phys. Rev. Lett. \textbf{123}, 122003 (2019).

\bibitem{BESIII:2021ee}
M.~Ablikim \textit{et al.} [BESIII],
Phys. Rev. D \textbf{104}, L091104 (2021).


\bibitem{BESIII:2021}
M.~Ablikim \textit{et al.} [BESIII],
Phys. Rev. D \textbf{104}, 052006 (2021).

\bibitem{BESIII:2022}
M.~Ablikim \textit{et al.} [BESIII],
[arXiv:2211.10755 [hep-ex]].

\bibitem{ALICE:2020}
S.~Acharya \textit{et al.} [ALICE],
Phys. Lett. B \textbf{802}, 135223 (2020).

\bibitem{ALICE:2022}
S.~Acharya \textit{et al.} [ALICE],
Phys. Lett. B \textbf{829}, 137060 (2022).

\bibitem{Li:2020}
H.~Li \textit{et al.} [GlueX],
AIP Conf. Proc. \textbf{2249}, 030037 (2020).

\bibitem{Pauli:2022}
P.~Pauli [GlueX],
EPJ Web Conf. \textbf{271}, 02001 (2022).

\bibitem{Zhou:2022}
X.~Zhou, L.~Yan, R.~B.~Ferroli and G.~Huang,
Symmetry \textbf{14}, 144 (2022).

\bibitem{Schonning:2023}
K.~Sch\"onning, V.~Batozskaya, P.~Adlarson and X.~Zhou,
[arXiv:2302.13071 [hep-ph]].

\bibitem{Baldini:2009}
R.~Baldini, S.~Pacetti, A.~Zallo and A.~Zichichi,
Eur. Phys. J. A \textbf{39}, 315 (2009).

\bibitem{Dalkarov:2010}
O.~D.~Dalkarov, P.~A.~Khakhulin and A.~Y.~Voronin,
Nucl. Phys. A \textbf{833}, 104 (2010). 

\bibitem{Haidenbauer:2016}
J.~Haidenbauer and U.-G.~Mei\ss{}ner,
Phys. Lett. B \textbf{761}, 456 (2016).

\bibitem{Yang:2018}
Y.~Yang and Z.~Lu,
Mod. Phys. Lett. A \textbf{33}, 1850133 (2018).

\bibitem{Cao:2018}
X.~Cao, J.-P.~Dai and Y.-P.~Xie,
Phys. Rev. D \textbf{98}, 094006 (2018).

\bibitem{Yang:2019}
Y.~Yang, D.-Y.~Chen and Z.~Lu,
Phys. Rev. D \textbf{100}, 073007 (2019).

\bibitem{Xiao:2019}
L.-Y.~Xiao, X.-Z.~Weng, X.-H.~Zhong and S.-L.~Zhu,
Chin. Phys. C \textbf{43}, 113105 (2019).

\bibitem{Ramalho:2020}
G.~Ramalho, M.~T.~Pe\~na and K.~Tsushima,
Phys. Rev. D \textbf{101}, 014014 (2020).

\bibitem{Haidenbauer:2021}
J.~Haidenbauer, U.-G.~Mei\ss{}ner and L.~Y.~Dai,
Phys. Rev. D \textbf{103}, 014028 (2021). 

\bibitem{Li:2022}
Z.~Y.~Li, A.~X.~Dai and J.~J.~Xie,
Chin. Phys. Lett. \textbf{39}, 011201 (2022).

\bibitem{Lin:2022}
Y.~H.~Lin, H.-W.~Hammer and U.-G.~Mei\ss{}ner,
Eur. Phys. J. C \textbf{82}, 1091 (2022).

\bibitem{Bystritskiy:2022}
Y.~M.~Bystritskiy and A.~I.~Ahmadov,
Phys. Rev. D \textbf{105}, 116012 (2022).

\bibitem{Tomasi-Gustafsson:2022}
E.~Tomasi-Gustafsson and S.~Pacetti,
Phys. Rev. C \textbf{106}, 035203 (2022).


\bibitem{Haidenbauer:1992}
J.~Haidenbauer, T.~Hippchen, K.~Holinde, B.~Holzenkamp, V.~Mull and J.~Speth,
Phys. Rev. C \textbf{45}, 931 (1992).

\bibitem{Haidenbauer:1992A}
  J.~Haidenbauer, K.~Holinde, V.~Mull and J.~Speth,
  Phys.\ Lett.\ B {\bf 291}, 223 (1992).

\bibitem{Haidenbauer:1992B}
  J.~Haidenbauer, K.~Holinde, V.~Mull and J.~Speth,
  Phys.\ Rev.\  C {\bf 46}, 2158 (1992).

\bibitem{Haidenbauer:1993}
  J.~Haidenbauer, K.~Holinde and J.~Speth,
  Nucl.\ Phys.\ A {\bf 562}, 317 (1993).


\bibitem{SibirtsevPRD}
  A.~Sibirtsev, J.~Haidenbauer, S.~Krewald, U.-G.~Mei\ss ner and A.~W.~Thomas,
  Phys.\ Rev.\ D {\bf 71}, 054010 (2005).

\bibitem{Kang:2015}
X.~W.~Kang, J.~Haidenbauer and U.-G.~Mei\ss{}ner,
Phys. Rev. D \textbf{91}, 074003 (2015).

\bibitem{Haidenbauer:2014}
J.~Haidenbauer, X.~W.~Kang and U.-G.~Mei\ss{}ner,
Nucl. Phys. A \textbf{929}, 102 (2014).

\bibitem{Gasparyan}
  A.~Gasparyan, J.~Haidenbauer and C.~Hanhart,
  Phys.\ Rev.\ C {\bf 72}, 034006 (2005).

\bibitem{Entem:2007}
D.~R.~Entem and F.~Fernandez,
Phys. Rev. D \textbf{75}, 014004 (2007).

 \bibitem{Hanhart}
  C.~Hanhart and K.~Nakayama,
  Phys.\ Lett.\ B {\bf 454}, 176 (1999)

\bibitem{Baru}
  V.~Baru, A.~M.~Gasparian, J.~Haidenbauer, A.~E.~Kudryavtsev and J.~Speth,
  Phys.\ Atom.\ Nucl.\  {\bf 64}, 579 (2001)
  [Yad.\ Fiz.\  {\bf 64}, 633 (2001)].

\bibitem{Watsonbook}
   M.~L.~Goldberger and K.~M.~Watson, {\it Collision Theory} (John Wiley and Sons, New York 1964), Chap.~9.3.

\bibitem{Carbonell:1993}
  J.~Carbonell, K.~V.~Protasov and O.~D.~Dalkarov,
  Phys.\ Lett.\ B {\bf 306}, 407 (1993).

\bibitem{Bugg:2004}
D.~V.~Bugg,
Eur. Phys. J. C \textbf{36}, 161 (2004). 

\bibitem{Milstein:2022}
A.~I.~Milstein and S.~G.~Salnikov,
Phys. Rev. D \textbf{105}, L031501 (2022). 

\bibitem{Cho:2017}
  S.~Cho {\it et al.} [ExHIC Collaboration],
  Prog.\ Part.\ Nucl.\ Phys.\  {\bf 95}, 279 (2017).

\bibitem{Haidenbauer:2018LL}
J.~Haidenbauer,
Nucl. Phys. A \textbf{981}, 1 (2019). 

\bibitem{Haidenbauer:2022LL}
J.~Haidenbauer and U.-G.~Mei\ss{}ner,
Phys. Lett. B \textbf{829}, 137074 (2022).

\end{thebibliography}
\end{document}